\documentclass{ifacconf}

\usepackage[utf8]{inputenc}
\usepackage{graphicx}           
\usepackage{natbib}             
\usepackage{amsmath}
\usepackage{amssymb}
\usepackage[short]{optidef}
\usepackage{xcolor}
\usepackage{caption}
\usepackage{pgfplots}
	\pgfplotsset{compat=newest}
	\usepgfplotslibrary{groupplots}
\usepackage{tikz}

\DeclareUnicodeCharacter{2212}{-}
\begin{document}
\begin{frontmatter}

\title{Combining system identification with reinforcement learning-based MPC} 

\author[]{Andreas B.\ Martinsen,}
\author[]{Anastasios M.\ Lekkas and}
\author[]{S{\'e}bastien\ Gros}

\vspace{0.3cm}

\begin{minipage}[c]{1.0\linewidth}
	\centering
	\it
	Department of Engineering Cybernetics,
	Norwegian University of Science and Technology (NTNU),
	O.\ S.\ Bragstads plass 2D,
	7491 Trondheim, Norway
	\\
	E-mails: \{andreas.b.martinsen,\allowbreak  anastasios.lekkas,\allowbreak sebastien.gros\}@ntnu.no
\end{minipage}

\begin{abstract} 
In this paper we propose and compare methods for combining system identification (SYSID) and reinforcement learning (RL) in the context of data-driven model predictive control (MPC). Assuming a known model structure of the controlled system, and considering a parametric MPC, the proposed approach simultaneously: a) Learns the parameters of the MPC using RL in order to optimize performance, and b) fits the observed model behaviour using SYSID. Six methods that avoid conflicts between the two optimization objectives are proposed and evaluated using a simple linear system. Based on the simulation results, hierarchical, parallel projection, nullspace projection, and singular value projection achieved the best performance.
\end{abstract}


\end{frontmatter}

\section{Introduction}
Reinforcement Learning (RL) is a powerful tool for tackling Markov Decision Processes (MDP) without depending on a model of the probability distributions underlying the state transitions. Most RL methods rely purely on observed state transitions, and realizations of the stage cost in order to increase the performance of the control policy. RL has drawn increasing attention due to recent high profile accomplishments made possible using function approximators \citep{busoniu2017reinforcement}.  Notable examples include performing at super human levels in games such as Go, chess and Atari \citep{silver2017mastering, silver2017masteringgo, mnih2013playing}, and robots learning to walk, fly without supervision, and perform complex manipulation \citep{wang2012machine, abbeel2007application, andrychowicz2018learning}. Most of these recent advances have been the result of RL with Deep Learning (DL) by using Deep Neural Networks (DNNs) as function approximators. While systems controlled by DNNs show a lot of promise, they are difficult to analyze, and in turn their behaviour is difficult to certify and trust.  

Model Predictive Control (MPC) is a popular approach for optimizing the closed loop performance of complex systems subject to constraints. MPC works by solving an optimal control problem at each control interval in order to find an optimal policy. The optimal control problem seeks to minimize the sum of stage costs over a horizon, provided a model of the system and the current observed state. While MPC is a well-studied approach, and an extensive literature exists on analysing its properties \citep{mayne2000constrained, rawlings2009optimizing}, the closed loop performance heavily relies on the accuracy of the underlying system model, which naturally presents challenges when significant unmodeled uncertainties are present. 

In recent works, such as \citep{gros2019data, zanon2019safe}, RL and MPC have been combined, by allowing RL to use a MPC as a function approximator. This approach allows to combine the benefits of data-driven optimization from RL with the tools available for analysing and certifying the closed loop performance of MPC. In this paper we extend the work by \cite{gros2019data}, by using a parametric MPC as a function approximator for performing RL, and combining it with on-line system identification (SYSID). The SYSID component is added with the purpose of aiding RL when there is a large model mismatch, as well as helping to improve the accuracy from the resulting MPC trajectory prediction. The main contribution of the paper are the methods for combining the competing optimization objectives of the RL and the SYSID in a way that minimizes plant model mismatch while not affecting the closed loop performance of the MPC. This paper focuses on the Q-learning approach to RL.

The paper is organized into five sections. Section 2 gives a brief overview of data-driven MPC, reinforcement learning and system identification. Section 3 describes several approaches for combining RL and SYSID in order to avoid loss in performance due to conflicting objectives. Section 4 shows simulation results for the different proposed methods, and finally, Section 5 concludes the paper. 

\section{Background}

\subsection{MPC as function approximator}
As in \cite{gros2019data}, we will use a parametric optimization problem as a function approximator for reinforcement learning. Given a stage cost $ L(\boldsymbol{x}, \boldsymbol{u})$ we can express the following MPC problem
\begin{mini!}{\boldsymbol{x}, \boldsymbol{u}, \boldsymbol{\sigma}}
    {\lambda_{\boldsymbol{\theta}}(\boldsymbol{x}_0) + 
    \sum_{i = 0}^{N - 1} \gamma^{i} \left( L(\boldsymbol{x}_i, \boldsymbol{u}_i) + L_{\boldsymbol{\theta}}(\boldsymbol{x}_i, \boldsymbol{u}_i) + \boldsymbol{\omega}^\top \boldsymbol{\sigma}_i \right) }{\label{eq:RLMPC}}{\notag}
    \breakObjective{\quad + \gamma^{N}V_{\boldsymbol{\theta}}^{f}(\boldsymbol{x}_N)}
    \addConstraint{\boldsymbol{x}_{i + 1} = f_{\boldsymbol{\theta}}(\boldsymbol{x}_i, \boldsymbol{u}_i)}
    \addConstraint{h(\boldsymbol{x}_i, \boldsymbol{u}_i) + h_{\boldsymbol{\theta}}(\boldsymbol{x}_i, \boldsymbol{u}_i) \leq \boldsymbol{\sigma}_i}
    \addConstraint{\boldsymbol{x}_0 = \boldsymbol{s}},
\end{mini!}
where we optimize the state, $\boldsymbol{x}$, action $\boldsymbol{u}$ and slack variables $\boldsymbol{\sigma}$ over the time horizon $N$. In the optimization problem, $\lambda_{\boldsymbol{\theta}}(\boldsymbol{x})$ is an initial cost modifier, $L(\boldsymbol{x}, \boldsymbol{u})$ is the stage cost, $L_{\boldsymbol{\theta}}(\boldsymbol{x}, \boldsymbol{u})$ is a parametric stage cost modifier, $V_{\boldsymbol{\theta}}^{f}(\boldsymbol{x})$ is a parametric terminal cost approximation, $f_{\boldsymbol{\theta}}(\boldsymbol{x}, \boldsymbol{u})$ is a parametric model approximation, $h(\boldsymbol{x}, \boldsymbol{u})$ and $h_{\boldsymbol{\theta}}(\boldsymbol{x}, \boldsymbol{u})$ are inequality constraints and inequality constraint modifiers, and $\gamma \in (0, 1]$ is the discount factor. The goal of the RL component is to modify the parameters $\boldsymbol{\theta}$ of the parametric optimization problem in order to find a policy $\pi_{\boldsymbol{\theta}}(\boldsymbol{x})$ that minimizes the expected cumulative discounted baseline stage cost:
\begin{equation*}
    \min_{\boldsymbol{\theta}} \mathbb{E} \left[ \sum_{i = 0}^{\infty} \gamma^{i} \bar{L}(\boldsymbol{x}_i, \pi_{\boldsymbol{\theta}} ({\boldsymbol{x}_i})) \right],
\end{equation*}
where the baseline stage cost $\bar{L}$ is defined as:
\begin{equation*}
    \bar{L}(\boldsymbol{x}_i, \boldsymbol{u}_i) = L(\boldsymbol{x}_i, \boldsymbol{u}_i) + \boldsymbol{\omega}^\top \max(0, h(\boldsymbol{x}_i, \boldsymbol{u}_i)).
\end{equation*}
Here the second term penalizes the constraint violations.

Ideally we would like strict constraints, however this would mean the MPC problem can become infeasible when model mismatch or disturbances cause constraint violations. In order to mitigate this problem, a slack penalty $\boldsymbol{\omega}$ is used, which is chosen large enough such that the constraints are only violated when the MPC becomes infeasible. For the RL, adding slack constraints is also important, as strict constraints means a penalty of $\infty$ for constraint violations, which most RL algorithms are not able to deal with.

\subsection{Value functions and policy}
Given the parametric optimization problem (\ref{eq:RLMPC}), we define the parametric action-value function as:
\begin{mini!}{\boldsymbol{x}, \boldsymbol{u}, \boldsymbol{\sigma}}
    {\text{(\ref{eq:RLMPC}a)}}{\label{eq:q_theta}}{Q_{\boldsymbol{\theta}}(\boldsymbol{s}, \boldsymbol{a}) =}
    \addConstraint{\text{(\ref{eq:RLMPC}b) - (\ref{eq:RLMPC}d)}}
    \addConstraint{\boldsymbol{u}_0 = \boldsymbol{a}},
\end{mini!}
which trivially satisfies the fundamental equalities underlying the Bellman equation:
\begin{equation} \label{eq:bellman_v}
    V_{\boldsymbol{\theta}}(\boldsymbol{s}) = \min_{\boldsymbol{a}} Q_{\boldsymbol{\theta}}(\boldsymbol{s}, \boldsymbol{a}), 
\end{equation}
\begin{equation} \label{eq:bellman_pi}
    \pi_{\boldsymbol{\theta}}(\boldsymbol{s}) = \text{arg}\min_{\boldsymbol{a}} Q_{\boldsymbol{\theta}}(\boldsymbol{s}, \boldsymbol{a}). 
\end{equation}

\subsection{Q-Learning}
A classical RL approach is Q-Learning \citep{watkins1989learning}. To perform Q-Learning for MPC we can use semi-gradient methods \citep{sutton2018reinforcement}, which are based on parameter updates driven by minimizing the temporal-difference error $\delta$: 
\begin{equation*}
    \delta_t = y_t -  Q_{\boldsymbol{\theta}}(\boldsymbol{s}_t, \boldsymbol{a}_t), 
\end{equation*}
where $y_t = \bar{L}(\boldsymbol{x}_t, \boldsymbol{u}_t) + \gamma  V_{\boldsymbol{\theta}}(\boldsymbol{x}_{t + 1})$ is the fixed target value. Defining the squared temporal-difference error as the minimization objective, and assuming that the target value is independent of the parameterization $\boldsymbol{\theta}$, we get the semi-gradient update:
\begin{equation}\label{eq:q-learning}
    \boldsymbol{\theta} \gets \boldsymbol{\theta} + \alpha \delta \nabla_{\boldsymbol{\theta}}Q_{\boldsymbol{\theta}}(\boldsymbol{x}_t, \boldsymbol{u}_t),
\end{equation}
where $\alpha > 0$ is the step-size or learning rate. For the classical semi-gradient Q-learning scheme given in (\ref{eq:q-learning}), a second order method can be implemented by using quasi-Newton steps instead of gradient steps. This results in the following update law: 
\begin{equation}\label{eq:q-learning_newton}
    \boldsymbol{\theta} \leftarrow \boldsymbol{\theta} + \alpha \delta  \boldsymbol{H}^{-1}\nabla_{\boldsymbol{\theta}}Q_{\boldsymbol{\theta}}(\boldsymbol{x}_t, \boldsymbol{u}_t),
\end{equation}
where $\boldsymbol{H} = \nabla_{\boldsymbol{\theta}}^2(y_t - Q_{\boldsymbol{\theta}}(\boldsymbol{x}_t, \boldsymbol{u}_t))^2$ is the Hessian of the error between the targets and the action-value function. For a batch of transitions, the problem becomes a nonlinear least squares problem: 
\begin{equation*}
    \min_{\boldsymbol{\theta}} \psi(\boldsymbol{\theta}),\quad\text{where}\quad \psi(\boldsymbol{\theta}) = \sum_t \delta_t^2
\end{equation*}
which may be solved using a Gauss-Newton method, as proposed in \cite{zanon2019practical}. The modified Gauss-Newton method gives the following update law:
\begin{equation}\label{eq:q-learning_gauss-newton_modified}
    \boldsymbol{\theta} \leftarrow \boldsymbol{\theta} + \alpha \underbrace{(\boldsymbol{J}_Q^{\top} \boldsymbol{J}_Q^{\phantom{\top}} + \lambda_Q \boldsymbol{I})^{-1} \boldsymbol{J}_Q^\top \boldsymbol{\delta}}_{:=\boldsymbol{\Delta \theta}_{Q}},
\end{equation}
where $\boldsymbol{J}_Q$ is the Jacobian of the action-value function over the batch in use, and $\boldsymbol{\delta}$ is the vector of temporal difference errors:
\begin{equation*}
    \boldsymbol{J}_Q = 
    \begin{bmatrix}
        \nabla_{\boldsymbol{\theta}} Q_{\boldsymbol{\theta}}(\boldsymbol{x}_{t, 1}, \boldsymbol{u}_{t, 1}) \\
        \nabla_{\boldsymbol{\theta}} Q_{\boldsymbol{\theta}}(\boldsymbol{x}_{t, 2}, \boldsymbol{u}_{t, 2}) \\
        \vdots \\
        \nabla_{\boldsymbol{\theta}} Q_{\boldsymbol{\theta}}(\boldsymbol{x}_{t, B}, \boldsymbol{u}_{t, B}) \\
    \end{bmatrix}, \quad
    \boldsymbol{\delta} = 
    \begin{bmatrix}
        \delta_{1} \\
        \delta_{2} \\
        \vdots\\
        \delta_{B} \\
    \end{bmatrix}
\end{equation*}
over the batch $\mathcal{B} = \{(\boldsymbol{x}_{t, i}, \boldsymbol{u}_{t, i}, \boldsymbol{x}_{t + 1, i}) | i \in 1 \hdots B\}$. The diagonal matrix $\lambda_Q \boldsymbol{I}$ is added such that $\boldsymbol{J}_Q^{\top} \boldsymbol{J}_Q + \lambda_Q \boldsymbol{I}$ is positive definite, and acts as a regularization of the Gauss-Newton method. 

It is worth noting that the semi-gradient Q-Learning method given above yields no guarantee to find the global optimum of the  parameter for nonlinear function approximators $Q_{\boldsymbol{\theta}}$. This limitation pertains to most applications of RL relying on nonlinear function approximators such as the commonly used DNN. It is also worth noting that in practice the parameterization $\boldsymbol{\theta}$ is limited. This means we in general are not able to fit the Q function globally, but rather that the formulation above fits the Q function to the distribution from which the samples are drawn.

\subsection{System Identification}
System identification offers a large set of tools for building mathematical models of dynamic systems, using measurements of the systems input and output signals. Based on the data-driven MPC scheme outlined in the previous section, we want an on-line parameter estimation method compatible with the parametric model. A classical SYSID approach is the Prediction Error Method (PEM) where the objective is to minimize the difference between the observed state and the predicted state given the observed transition $(\boldsymbol{x}_t, \boldsymbol{u}_t, \boldsymbol{x}_{t + 1})$. For a parametric model approximation of the form:
\begin{equation*}
    \hat{\boldsymbol{x}}_{t + 1} = f_{\boldsymbol{\theta}}(\boldsymbol{x}_t, \boldsymbol{u}_t),
\end{equation*}
the state error $\boldsymbol{e}$ between the parametric model and the observed state can then be expressed as follows:
\begin{equation*}
    \boldsymbol{e}_t = \boldsymbol{x}_{t + 1} - \hat{\boldsymbol{x}}_{t + 1} = \boldsymbol{x}_{t + 1} - f_{\boldsymbol{\theta}}(\boldsymbol{x}_t, \boldsymbol{u}_t).
\end{equation*}
In the simplest case, where the state vector $\boldsymbol{x}$ is fully observable, PEM can be performed by minimizing the squared error between the observed state, and the predicted state:
\begin{equation*}
    \min_{\boldsymbol{\theta}} \phi(\boldsymbol{\theta}) ,\quad \text{where}\quad      \phi(\boldsymbol{\theta}) = \boldsymbol{e}^\top \boldsymbol{e}
\end{equation*}
where $\boldsymbol{e}$ collects a batch of measurements $\boldsymbol{e}_i$. This optimization problem can be tackled via gradient descent, giving the following update law: 
\begin{equation*}
   \boldsymbol{\theta} \gets \boldsymbol{\theta} - \beta \nabla_{\boldsymbol{\theta}} \boldsymbol{e}^\top \boldsymbol{e},
\end{equation*}
where $\beta$ is the learning rate. Since $\boldsymbol{\theta}$ are all the parameters appearing in the MPC (\ref{eq:RLMPC}), PEM is in practise only modifying the subset of the parameters $\boldsymbol{\theta}$ that appear in the parametric model. For faster learning, we propose using a second order approach, and perform quasi-Newton steps on the parameters. One such method is the modified Gauss-Newton method, which for a batch of transitions reads as follows: 
\begin{equation}\label{eq:sysid_gauss-newton_modified}
    \boldsymbol{\theta} \leftarrow \boldsymbol{\theta} + \beta \underbrace{(\boldsymbol{J}_f^{\top} \boldsymbol{J}_f^{\phantom{\top}} + \lambda_f \boldsymbol{I})^{-1} \boldsymbol{J}_f^\top \boldsymbol{e}}_{:=\boldsymbol{\Delta \theta}_{f}},
\end{equation}
where $\boldsymbol{J}_f$ is the Jacobian of parametric system model, and $\boldsymbol{e}$ is the vector of model errors over a batch $\mathcal{B} = \{(\boldsymbol{x}_{t, i}, \boldsymbol{u}_{t, i}, \boldsymbol{x}_{t + 1, i}) | i \in 1 \hdots B\}$. 
\begin{equation*}
    \boldsymbol{J}_f = 
    \begin{bmatrix}
        \nabla_{\boldsymbol{\theta}} f_{\boldsymbol{\theta}}(\boldsymbol{x}_{t, 1}, \boldsymbol{u}_{t, 1}) \\
        \nabla_{\boldsymbol{\theta}} f_{\boldsymbol{\theta}}(\boldsymbol{x}_{t, 2}, \boldsymbol{u}_{t, 2}) \\
        \vdots \\
        \nabla_{\boldsymbol{\theta}} f_{\boldsymbol{\theta}}(\boldsymbol{x}_{t, B}, \boldsymbol{u}_{t, B}) \\
    \end{bmatrix}, \:
    \boldsymbol{e} = 
    \begin{bmatrix}
        \boldsymbol{x}_{t + 1, 1} - f_{\boldsymbol{\theta}}(\boldsymbol{x}_{t, 1}, \boldsymbol{u}_{t, 1}) \\
        \boldsymbol{x}_{t + 1, 2} - f_{\boldsymbol{\theta}}(\boldsymbol{x}_{t, 2}, \boldsymbol{u}_{t, 2}) \\
        \vdots\\
        \boldsymbol{x}_{t + 1, B} - f_{\boldsymbol{\theta}}(\boldsymbol{x}_{t, B}, \boldsymbol{u}_{t, B}) \\
    \end{bmatrix}
\end{equation*}
Similarly to RL, for a batch of transitions, the problem becomes a least-squares problem, fitting the observed transitions to the parametric model. 

It is worth noting that for a linear parameterization, the Gauss-Newton method gives convergence to the least squares solution over the batch in one step, when $\beta = 1$ and $\lambda_f = 0$. It is also worth noting that since PEM only works on a subset of the parameters $\boldsymbol{\theta}$, $\boldsymbol{J}_f$ is rank deficient and hence $\boldsymbol{J}_f^{\top} \boldsymbol{J}_f^{\phantom{\top}}$ is singular by construction. Choosing the regularization term $\lambda_f > 0$, will ensure $\boldsymbol{J}_f^{\top} \boldsymbol{J}_f^{\phantom{\top}} + \lambda_f \boldsymbol{I}$ is nonsingular and hence invertible. For the regularization parameter $\lambda_f$ and $\lambda_Q$ we typically want to choose a small value, to get performance close to the pure Gauss-Newton method, while only slightly regularizing in order to avoid issues arising form a singular Hessian approximation. 

\section{System identification for data-driven MPC}
The Prediction Error Method and Reinforcement Learning are modifying the same parameter vector $\boldsymbol{\theta}$, but operate using two different objectives. RL is targeting policy optimization by minimizing the temporal difference error against a fixed target, while PEM is fitting the parametric model to the observed state transitions.

A combination of the two methods then becomes a multi-objective optimization problem. The simplest approach is to directly combine the steps from both the Q-Learning and PEM. Using the second order update laws in (\ref{eq:q-learning_gauss-newton_modified}) and (\ref{eq:sysid_gauss-newton_modified}), with the parameter updates $\boldsymbol{\Delta \theta}_{Q}$ and $\boldsymbol{\Delta \theta}_{f}$ respectively, we get the following: 
\begin{equation}\label{eq:sum_of_grads}
    \boldsymbol{\theta} \leftarrow \boldsymbol{\theta} + \alpha \boldsymbol{\Delta \theta}_{Q} + \beta \boldsymbol{\Delta \theta}_{f}
\end{equation}
Here the step-lengths $\alpha$ and $\beta$ can be thought of as the weighting between the Q-Learning and SYSID respectively. However, the end goal is arguably to maximize the closed-loop performance of the MPC scheme rather than minimizing the prediction error of the model, hence if the two objectives are competing, the RL objective should be prioritized. This suggests that an alternative to the naive sum of update laws approach must be considered. 

\subsection{Hierarchical multi-objective approach}
In order to introduce a hierarchy between minimizing the PEM and RL objectives, we can consider the optimization problem: 
\begin{subequations}
\label{eq:Formal:Hierarchic}
\begin{align}
\min_{\boldsymbol{\theta}}&\quad \phi(\boldsymbol{\theta}), \\
\mathrm{s.t.}&\quad \nabla_{\boldsymbol{\theta}} \psi(\boldsymbol{\theta}) = 0,
\end{align}
\end{subequations}
which requires $\boldsymbol\theta$ to minimize the PEM while being a stationary point of the RL objsective. If $\nabla^2_{\boldsymbol{\theta}} \psi(\boldsymbol{\theta}) \geq 0$ is satisfied at the solution of \eqref{eq:Formal:Hierarchic}, then $\boldsymbol\theta$ is a (local) minimizer of the RL objective. The KKT conditions associated to \eqref{eq:Formal:Hierarchic} read as:
\begin{subequations}
\label{eq:Formal:Hierarchic:KKT}
\begin{align}
\nabla_{\boldsymbol{\theta}}\phi(\boldsymbol{\theta}) + \nabla_{\boldsymbol{\theta}}^2 \psi(\boldsymbol{\theta}) \boldsymbol{\lambda} &=0, \\
\nabla_{\boldsymbol{\theta}} \psi(\boldsymbol{\theta}) &= 0.
\end{align}
\end{subequations}
A quasi-Newton step on \eqref{eq:Formal:Hierarchic:KKT} reads as:
\begin{subequations}
\label{eq:Formal:Hierarchic:Step}
\begin{align}
\nabla_{\boldsymbol{\theta}}^2\phi(\boldsymbol{\theta})\Delta\boldsymbol \theta + \nabla_{\boldsymbol{\theta}}^2 \psi(\boldsymbol{\theta}) \boldsymbol{\lambda} &= - \nabla_{\boldsymbol{\theta}}\phi(\boldsymbol{\theta}), \\
\nabla^2_{\boldsymbol{\theta}} \psi(\boldsymbol{\theta})\Delta\boldsymbol{\theta} &= -\nabla_{\boldsymbol{\theta}} \psi(\boldsymbol{\theta}).
\end{align}
\end{subequations}
Let us consider a (possibly $\boldsymbol\theta$-dependent) nullspace / full-space decomposition of the RL Hessian $\nabla^2_{\boldsymbol{\theta}} \psi$, i.e. $\boldsymbol{\mathcal{N}},\boldsymbol{\mathcal{F}}$ such that:
\begin{align}
\nabla^2_{\boldsymbol{\theta}} \psi(\boldsymbol{\theta})\boldsymbol{\mathcal{N}} = 0,\quad \left[\boldsymbol{\mathcal{N}}\,\,\, \boldsymbol{\mathcal{F}}\right]\,\, \text{full rank},\quad \boldsymbol{\mathcal{N}}^\top \boldsymbol{\mathcal{F}} = 0,
\end{align}
and the associated decomposition of the primal step $\Delta\boldsymbol{\theta}$:
\begin{align}
\Delta\boldsymbol{\theta} = \boldsymbol{\mathcal{N}} \boldsymbol n + \boldsymbol{\mathcal{F}} \boldsymbol f.
\end{align}
We then observe that the primal quasi-Newton step given by \eqref{eq:Formal:Hierarchic:Step} can be decomposed into:
\begin{subequations}
\label{eq:Formal:Hierarchic:Step:Decomposition}
\begin{align}
\boldsymbol{\mathcal{N}}^\top \nabla_{\boldsymbol{\theta}}^2\phi  \boldsymbol{\mathcal{N}} \boldsymbol n + \boldsymbol{\mathcal{N}}^\top \nabla_{\boldsymbol{\theta}}^2\phi  \boldsymbol{\mathcal{F}} \boldsymbol f  &= -\boldsymbol{\mathcal{N}}^\top \nabla_{\boldsymbol{\theta}}\phi, \\
\boldsymbol{\mathcal{F}}^\top\nabla^2_{\boldsymbol{\theta}} \psi\boldsymbol{\mathcal{F}} \boldsymbol f &= -\boldsymbol{\mathcal{F}}^\top\nabla_{\boldsymbol{\theta}} \psi.
\end{align}
\end{subequations}
One can then verify that:
\begin{subequations}
\label{eq:Formal:Hierarchic:Step:Decomposition:Solve}
\begin{align}
 \boldsymbol n   &= -\left(\boldsymbol{\mathcal{N}}^\top \nabla_{\boldsymbol{\theta}}^2\phi  \boldsymbol{\mathcal{N}}\right)^\dagger\left(\boldsymbol{\mathcal{N}}^\top \nabla_{\boldsymbol{\theta}}\phi - \boldsymbol{\mathcal{N}}^\top \nabla_{\boldsymbol{\theta}}^2\phi  \boldsymbol{\mathcal{F}} \boldsymbol f\right), \\
 \boldsymbol f &= -\left(\boldsymbol{\mathcal{F}}^\top\nabla^2_{\boldsymbol{\theta}} \psi\boldsymbol{\mathcal{F}}\right)^{-1}\boldsymbol{\mathcal{F}}^\top\nabla_{\boldsymbol{\theta}} \psi,
\end{align}
\end{subequations}
where $.^\dagger$ stands for the Moore-Penrose pseudo-inverse. Let us label:
\begin{align}
\label{eq:PINV:PEM}
\nabla_{\boldsymbol{\theta}}^2\phi^\dagger_\perp = \boldsymbol{\mathcal{N}}\left(\boldsymbol{\mathcal{N}}^\top \nabla_{\boldsymbol{\theta}}^2\phi  \boldsymbol{\mathcal{N}}\right)^\dagger\boldsymbol{\mathcal{N}}^\top,
\end{align}
the pseudo-inverse of the SYSID Hessian $\nabla_{\boldsymbol{\theta}}^2\phi$ projected in the nullpsace of the RL Hessian. Let us additionally label
\begin{align}
\label{eq:Label:Decomposition}
\Delta \boldsymbol \theta_Q^H = \boldsymbol{\mathcal{F}}\boldsymbol f,\quad
\Delta \boldsymbol \theta_f^H  = \boldsymbol{\mathcal{N}}\boldsymbol n.
\end{align}
The primal step $\Delta \boldsymbol \theta$ then reads as:
\begin{subequations}
\label{eq:Formal:Hierarchic:Step:Decomposition:Final}
\begin{align}
\Delta \boldsymbol \theta &=   \Delta \boldsymbol \theta_Q^H + \Delta \boldsymbol \theta_f^H, \\
\Delta \boldsymbol \theta_Q^H &=-\boldsymbol{\mathcal{F}}^\top\left(\boldsymbol{\mathcal{F}}^\top\nabla^2_{\boldsymbol{\theta}} \psi\boldsymbol{\mathcal{F}}\right)^{-1}\boldsymbol{\mathcal{F}}^\top\nabla_{\boldsymbol{\theta}} \psi \label{eq:Formal:Hierarchic:Step:Decomposition:dQ} = -\nabla^2_{\boldsymbol{\theta}} \psi^\dagger \nabla_{\boldsymbol{\theta}} \psi, \\
\Delta \boldsymbol \theta_f^H  &= -\nabla_{\boldsymbol{\theta}}^2\phi^\dagger_\perp\mathcal  \nabla_{\boldsymbol{\theta}}\phi  +\nabla_{\boldsymbol{\theta}}^2\phi^\dagger_\perp \nabla_{\boldsymbol{\theta}}^2\phi \Delta \boldsymbol \theta_Q^H.  \label{eq:Formal:Hierarchic:Step:Decomposition:dPEM} 
\end{align}
\end{subequations}
In practice, pseudo-inverses are not always numerically stable. In order to alleviate this potential issue, we can use regularizations of the PEM and RL Hessians instead, i.e. we can select $\lambda_{Q},\, \lambda_{f} > 0$ and use:
\begin{subequations}
\label{eq:Proj:Regul}
\begin{align}
    \Delta \boldsymbol \theta_Q^H &=-\left(\nabla^2_{\boldsymbol{\theta}} \psi+\lambda_f \boldsymbol{I} \right)^{-1}\nabla_{\boldsymbol{\theta}} \psi, \\
\nabla_{\boldsymbol{\theta}}^2\phi^\dagger_\perp &= \boldsymbol{\mathcal{N}}\left(\boldsymbol{\mathcal{N}}^\top \left(\nabla_{\boldsymbol{\theta}}^2\phi  +\lambda_Q \boldsymbol{I}\right)\boldsymbol{\mathcal{N}}\right)^{-1}\boldsymbol{\mathcal{N}}^\top,
\end{align}
\end{subequations}
together with \eqref{eq:Formal:Hierarchic:Step:Decomposition:dPEM} instead of \eqref{eq:Formal:Hierarchic:Step:Decomposition:dQ} and \eqref{eq:PINV:PEM}. For $\lambda_{Q,f}\rightarrow 0$, \eqref{eq:Proj:Regul}-\eqref{eq:Formal:Hierarchic:Step:Decomposition:dPEM} asymptotically deliver the same steps as \eqref{eq:Formal:Hierarchic:Step:Decomposition:Final}.
\subsection{Projected steps}
In this section we will discuss several projections we can perform in order to mitigate conflicts between the two optimization objectives. As discussed earlier, we typically want the RL updates to dominate, as these are directly related to the MPC closed-loop performance.

\subsubsection{Parallel projection}
We first consider a parallel projection, where the PEM step $\boldsymbol{\Delta \theta}_{f}$ is projected along the RL step $\boldsymbol{\Delta \theta}_{Q}$, giving the following projected PEM step
\begin{equation*}
  \boldsymbol{\Delta \theta}_{f}^\parallel =   \frac{\boldsymbol{\Delta \theta}_{Q}^{\phantom{\top}} \boldsymbol{\Delta \theta}_{Q}^\top}{\boldsymbol{\Delta \theta}_{Q}^\top \boldsymbol{\Delta \theta}_{Q}^{\phantom{\top}}} \boldsymbol{\Delta \theta}_{f}
\end{equation*}
Intuitively, this projection can be thought of as an adaptive step-length for the RL step, i.e. the SYSID modifies the RL step-length in the direction that improves the SYSID loss.

\subsubsection{Orthogonal projection}
Similar to the parallel projection, we may use the orthogonal projection:
\begin{equation*}
  \boldsymbol{\Delta \theta}_{f}^\perp  =    \left( \boldsymbol{I} - \frac{\boldsymbol{\Delta \theta}_{Q}^{\phantom{\top}} \boldsymbol{\Delta \theta}_{Q}^\top}{\boldsymbol{\Delta \theta}_{Q}^\top \boldsymbol{\Delta \theta}_{Q}^{\phantom{\top}}} \right) \boldsymbol{\Delta \theta}_{f}
\end{equation*}
The orthogonal projection is dual to the parallel projection, in that it does not effect the length of the of the RL step. It may however have the drawback of working against the RL step since we do not account for the sensitivity of the RL objective in the orthogonal direction. This can be easily be seen in the case where a optimum of the RL objective is achieved, i.e. $\delta \nabla_{\boldsymbol{\theta}}Q_{\boldsymbol{\theta}}(\boldsymbol{x}, \boldsymbol{u}) = \boldsymbol{0}$, where any PEM step will in general push the parameters away form the RL optimum.

\subsubsection{Nullspace projection}
Based on the heriarcical optimization problem in (\ref{eq:Formal:Hierarchic:Step:Decomposition:dPEM}), we see that in the particular case that $\nabla_{\boldsymbol{\theta}}^2\phi = c^{-1}\cdot \boldsymbol{I}$ holds, the hierarchical optimization problem reduces to simply projecting the PEM step into the nullspace of the RL step:
\begin{subequations}
\begin{align}
\label{eq:Formal:Hierarchic:Step:Decomposition:Reduced}
\nabla_{\boldsymbol{\theta}}^2\phi^\dagger_\perp &= c \boldsymbol{\mathcal{N}}\boldsymbol{\mathcal{N}}^\top,\\
\Delta \boldsymbol \theta_f^H  &= -c \boldsymbol{\mathcal{N}}\boldsymbol{\mathcal{N}}^\top \nabla_{\boldsymbol{\theta}}\phi = -\boldsymbol{\mathcal{N}}\boldsymbol{\mathcal{N}}^\top \nabla_{\boldsymbol{\theta}}^2\phi^{-1}\nabla_{\boldsymbol{\theta}}\phi.
\end{align}
\end{subequations}
Using this nullspace projection, with the gauss newton approach in (\ref{eq:sysid_gauss-newton_modified}) we get the following update law:  
\begin{equation*}
   \boldsymbol{\Delta \theta}_{f}^\mathrm N  =    \boldsymbol{\mathcal{N}}\boldsymbol{\mathcal{N}}^\top \boldsymbol{\Delta \theta}_{f}.
\end{equation*}
Choosing this simplified nullspace projection, the PEM step is projected into a direction for which the value function is not sensitive, hence the gradient step for the SYSID will not effect the primary goal of optimizing the RL objective. The nullspace projection may also be thought of as a regularization of the RL objective.

\subsubsection{Smallest singular value projection}
The nullspace projects the PEM steps into the nullspace of $\boldsymbol H$, i.e. the space where the singular values of $\boldsymbol H$ are zero. As a generalization of the nullspace projection, we can project the PEM steps into the space where the Hessian is the least sensitive. Using the singular value decomposition of the Hessian of the temporal difference loss.
\begin{equation*}
    \boldsymbol{U} \boldsymbol{\Sigma} \boldsymbol{V} = \boldsymbol{H}
\end{equation*}
We can extract an orthonomal basis of the $p$ smallest singular values $\underline{\boldsymbol{V}}$ as the last $p$ rows of $\boldsymbol{V}$. The projection into the $p$ smallest singular values is then given by the following.
\begin{equation*}
    \boldsymbol{\Delta \theta}_{f}^{S} = \underline{\boldsymbol{V}}^\top \underline{\boldsymbol{V}} \boldsymbol{\Delta \theta}_{f}
\end{equation*}
We can alternatively choose $p$ to be the number of singular values under a certain threshold. Note that if we choose $p$ to be the number of singular values equal to zero, the projection becomes equivalent to the nullspace projection. While the nullspace projection will give no progress if $\boldsymbol{H}$ is full rank, the singular value projection ensures some progress on the PEM objective, at a small cost to to the RL objective. 

\section{Simulations}
\begin{figure}
    \centering
    %
    \input{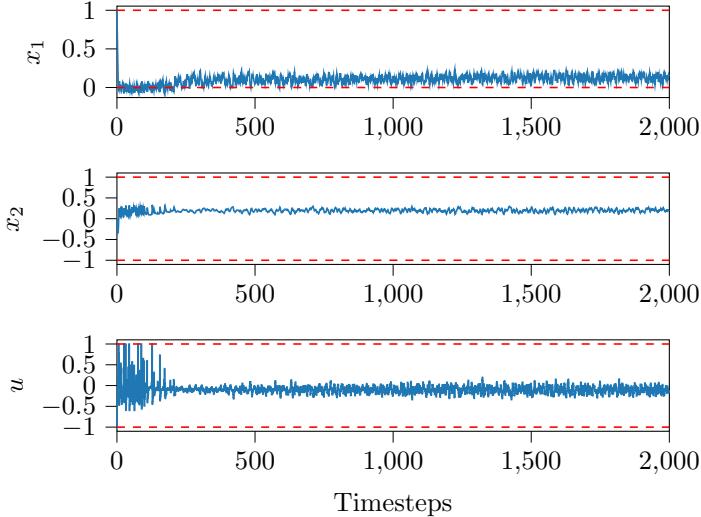}%
    \vspace*{-8mm}

    \caption{Baseline when using only reinforcement learning (\ref{eq:q-learning_gauss-newton_modified}). The optimal unconstrained solution would be to regulate the system to $x_1 = 0$, however due to the constraint, and disturbances this is no longer the case.}
    \label{fig:baseline}
\end{figure}

In this section we will compare the performance of the different RL MPC modifications proposed above. In order to gauge the results we consider the following simple linear MPC problem: 
\begin{mini!}
    {\boldsymbol{x}, \boldsymbol{u}, \boldsymbol{\sigma}}
    {\sum_{i = 0}^{N - 1} \gamma^{i} \left( ||\boldsymbol{x}_i||^2 + \frac{1}{2}||\boldsymbol{u}_i||^2 + \boldsymbol{f}^{\top} \begin{bmatrix} \boldsymbol{x}_i \\ \boldsymbol{u}_i \end{bmatrix} + \boldsymbol{\omega}^\top \boldsymbol{\sigma}_i \right)}{}{\notag}
    \breakObjective{\quad + V_0 +  \gamma^{N} \boldsymbol{x}_N^\top \boldsymbol{S} \boldsymbol{x}_N}
    \addConstraint{\boldsymbol{x}_{i + 1} = \boldsymbol{A}\boldsymbol{x}_i + \boldsymbol{B}\boldsymbol{u}_i + \boldsymbol{b}}
    \addConstraint{\begin{bmatrix} 0\\ - 1 \end{bmatrix} + \underline{\boldsymbol{x}} - \boldsymbol{\sigma}_i \leq \boldsymbol{x}_i \leq \begin{bmatrix} 1\\ 1 \end{bmatrix} + \bar{\boldsymbol{x}} + \boldsymbol{\sigma}_i}
    \addConstraint{-1 \leq \boldsymbol{u}_i \leq 1}
\end{mini!}
where the parameters $\boldsymbol{\theta}$ of the optimization problem are given as:
\begin{equation*}
    \boldsymbol{\theta} = (V_0, \boldsymbol{f}, \boldsymbol{S}, \boldsymbol{A}, \boldsymbol{B}, \boldsymbol{b}, \underline{\boldsymbol{x}}, \bar{\boldsymbol{x}})
\end{equation*}
For the initial model parameter guess used in the MPC we have the following:
\begin{equation*}
    \boldsymbol{A} = 
    \begin{bmatrix} 
        1.0 & 0.25 \\ 
        0.0 & 1.0
    \end{bmatrix}
    , \quad 
    \boldsymbol{B} =
    \begin{bmatrix} 
        0.0312 \\ 
        0.25  
    \end{bmatrix}
    , \quad
    \boldsymbol{b} = 
    \begin{bmatrix} 
        0 \\ 
        0  
    \end{bmatrix}
\end{equation*}
Additionally the terminal cost matrix $\boldsymbol{S}$ was chosen as the solution to the discrete-time algebraic Riccati equation, while the rest of the parameters were initialized to zero. For the real process we used the following dynamics: 
\begin{equation*}
    \boldsymbol{x}_{i + 1} = 
    \begin{bmatrix} 
        0.9 & 0.35 \\ 
        0.0 & 1.1
    \end{bmatrix}
    \boldsymbol{x}_i + 
    \begin{bmatrix} 
        0.0813 \\ 
        0.2  
    \end{bmatrix}
    \boldsymbol{u}_i + 
    \begin{bmatrix} 
        e_k \\ 
        0  
    \end{bmatrix}
\end{equation*}
where $e_k$ is uniformly distributed on the interval $[-0.1, 0]$. The disturbance will have the effect of pushing the first state towards the lower bound such that the constraint is violated, and in turn incurring a large cost. To prevent this from happening and perform optimally, the RL algorithm must modify the parameters $\boldsymbol{\theta}$. In Figure \ref{fig:baseline} the states $\boldsymbol{x}$ and action $\boldsymbol{u}$ are shown for the baseline method, which only uses pure RL steps. As seen in the figure, the constraints on the first state $x_1$ are violated in the beginning, but by updating the parameters using RL, the system quickly learns to avoid the constraints, while at the same time staying as close to them as possible in order to minimize the discounted stage cost.

\begin{figure}
    \centering
    %
    \input{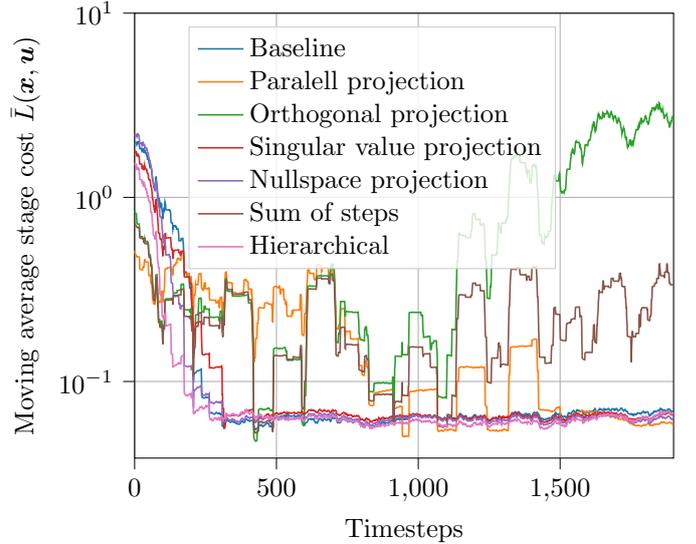}%
    \vspace*{-8mm}

    \caption{Moving average stage cost over $100$ steps. Jumps/steps in performance indicates constraint violations, which results in a large cost.}
    \label{fig:stage_cost}
\end{figure}

\begin{figure}
    \centering
    %
    \input{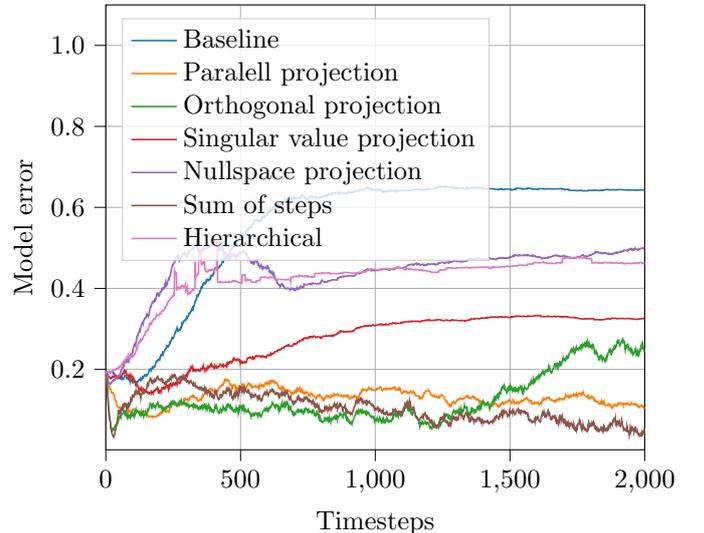}%
    \vspace*{-8mm}

    \caption{Norm of the parameter error for the model parameters $\boldsymbol{A}$, $\boldsymbol{B}$ and $\boldsymbol{b}$. Lower error means the parametric model in the MPC is closer to the simulated model.}
    \label{fig:parameter_error}
\end{figure}

\begin{figure}
    \centering
    %
    \input{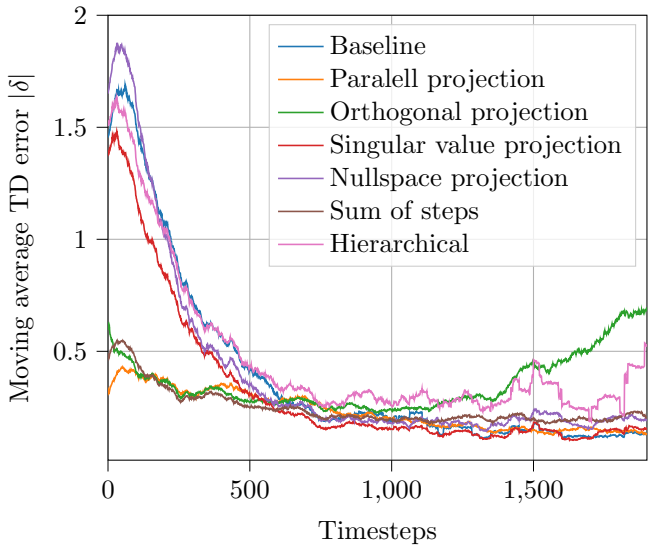}%
    \vspace*{-8mm}

    \vspace*{4mm}
    \caption{Moving average absolute temporal difference error $|\delta|$ over 100 steps.}
    \label{fig:temporal_difference_eror}
\end{figure}

\begin{figure}
    \centering
    %
    \input{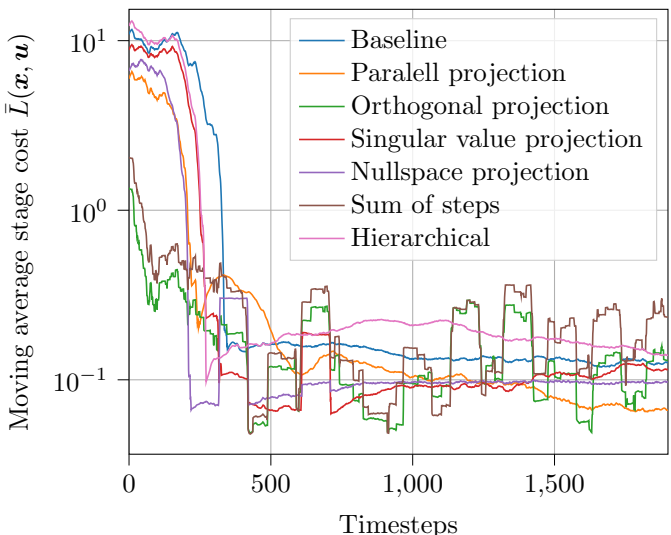}%
    \vspace*{-8mm}

    \caption{Moving average stage cost over $100$ steps using poor initial model parameters, we see a clear improvement in performance in the early learning stage.}
    \label{fig:stage_cost_poor_initial}
\end{figure}

\begin{figure}
    \centering
    %
    \input{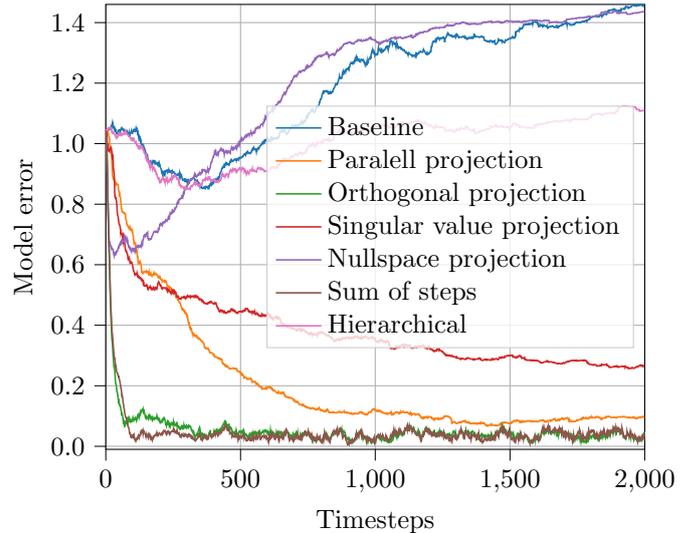}%
    \vspace*{-8mm}

    \caption{Norm of the parameter error for the parametric model with poor initial model parameters.}
    \label{fig:parameter_error_poor_initial}
\end{figure}

Running the on-line RL together with the proposed PEM methods, we get the results seen in Figures \ref{fig:stage_cost}, \ref{fig:parameter_error} and \ref{fig:temporal_difference_eror}. Figure \ref{fig:stage_cost} shows the moving average stage cost, which is a good performance measure of the closed loop performance of the MPC. From the results we see the the hierarchical, parallel, singular value and nullspace projections all converge to a slightly better performance than the baseline, while the orthogonal projection, and weighted sum of steps perform worse then the baseline. The drop in performance of the orthogonal projection, weighted sum of steps, and to a certain degree the parallel projection, is the result of competing objectives. This is also reflected in the parameter error as seen in Figure \ref{fig:parameter_error}, where the model fit comes at the expense of closed loop performance. Looking at the temporal difference error in Figure \ref{fig:temporal_difference_eror}, we see that most of the proposed methods give  faster initial convergence. This is a result of the improved plant model mismatch which in turn gives better value function estimates from the MPC. A similar observation can be made in Figure \ref{fig:stage_cost_poor_initial} and \ref{fig:parameter_error_poor_initial}, where the initial model parameters were chosen as a double integrator:
\begin{equation*}
    \boldsymbol{A} = 
    \begin{bmatrix} 
        1 & 1 \\ 
        0 & 1
    \end{bmatrix}
    , \quad 
    \boldsymbol{B} =
    \begin{bmatrix} 
        0 \\ 
        1  
    \end{bmatrix}
    , \quad
    \boldsymbol{b} = 
    \begin{bmatrix} 
        0 \\ 
        0  
    \end{bmatrix},
\end{equation*}
giving a lager plant model mismatch. From the results we see that all the proposed methods have a better initial convergence of the closed loop performance, with the parallel, singular value and nullspace projections, also giving better final closed loop performance. For the parameter error, we see a significant improvement of all methods, except for the hierarchical and nullspace projection, in comparison with the baseline. The results are in line with the constraints imposed by the different projections, where the hierarchical and nullspace projection being the most conservative, and the summation of gradients being the least conservative. 
 
\section{Conclusion}
In this paper we proposed and tested a number of strategies for combing RL, PEM and data-driven MPC in order to perform on-line learning and control. The main contribution is the addition of PEM as an on-line system identification method, which is added in order to aid the RL when there is a large plant model mismatch, as well as help to get better accuracy from the resulting MPC trajectory prediction. The proposed parallel, singular value and nullspace projection methods show promising results in terms of decreasing plant model miss-match, and giving slightly better closed loop MPC performance than using pure RL, while the orthogonal projection, and sum of steps resulted in improved model fit, however at the cost of closed loop performance of the proposed MPC scheme. In conclusion, combing PEM with RL, can give better initial learning when we do not have a good initial guess for the parameters, as well as lead to better overall performance of the closed loop MPC, without significant additional computational overhead.

For future work, it is of interest to look at methods for adaptively changing the step-length of the two objectives. For example choosing a step-length $\beta$ dependant on the RL step, may help mitigate the problem of competing objectives, and in turn improve the performance of the proposed methods. Combining the proposed method with policy gradient, is also an area of interest, as policy gradient methods offer a way of directly optimizing the policy.


\bibliography{ifacconf}             

\appendix

\end{document}